\documentclass[10pt, conference, compsocconf]{IEEEtran}

\usepackage[utf8]{inputenc}
\usepackage[english]{babel}
\usepackage[dvipsnames]{xcolor}
\usepackage{cite}
\usepackage{graphicx, tikz, epstopdf, adjustbox, stfloats}
\usepackage{balance}
\usepackage[hyphens]{url}
\usepackage[inline]{enumitem}
\usepackage[hidelinks]{hyperref}
\usepackage{todonotes}
\usepackage{subcaption}
\graphicspath{{figs/}}

\usepackage[]{cleveref} 
\crefformat{section}{Section~#2#1#3}
\crefformat{figure}{Fig.~#2#1#3}

\usepackage[final]{microtype}
\makeatletter
\newcommand\subparagraph{%
    \@startsection{subparagraph}{5}
    {\parindent}
    {3.25ex \@plus 1ex \@minus .2ex}
    {-1em}
    {\normalfont\normalsize\bfseries}}
\makeatother
\usepackage[compact]{titlesec}
\let\subparagraph\relax 
\titlespacing\section{0pt}{6pt plus 4pt minus 2pt}{2pt plus 2pt minus 2pt}
\titlespacing{\subsection}{0pt}{4pt plus 2pt minus 1pt}{2pt plus 1pt minus 1pt}
\titlespacing{\subsubsection}{0pt}{4pt plus 2pt minus 1pt}{2pt plus 1pt minus 1pt}
\usepackage{etoolbox}
\makeatletter
\patchcmd{\ttlh@hang}{\parindent\z@}{\parindent\z@\leavevmode}{}{}
\patchcmd{\ttlh@hang}{\noindent}{}{}{}
\makeatother

\usepackage{multirow}
\usepackage[none]{hyphenat}

\usepackage{xcolor}

\newcommand*\circled[1]{\tikz[baseline=(char.base)]{
        \node[shape=circle,draw,inner sep=2pt, Maroon, fill=Maroon] (char)
               {\color{white}\scriptsize\textbf{#1}};}%
        }

\begin{document}
%
\title{Kayotee: A Fault Injection-based System to Assess  the Safety and Reliability of Autonomous Vehicles to  Faults and Errors\\
}
\author{\IEEEauthorblockN{
    Saurabh Jha\IEEEauthorrefmark{1,2},
    Timothy Tsai\IEEEauthorrefmark{2},
    Siva Hari\IEEEauthorrefmark{2},
    Michael Sullivan\IEEEauthorrefmark{2},
    Zbigniew Kalbarczyk\IEEEauthorrefmark{1}, \\
    Stephen W. Keckler\IEEEauthorrefmark{2},
    and
    Ravishankar K. Iyer\IEEEauthorrefmark{1}
    }
    \IEEEauthorblockA{\IEEEauthorrefmark{1}University of Illinois at Urbana-Champaign, Urbana-Champaign, IL, USA, 61801}
    \IEEEauthorblockA{\IEEEauthorrefmark{2}Nvidia Corporation,  Santa Clara, CA, USA, 94086}
}

\maketitle

  \begin{abstract}
      Fully autonomous vehicles (AVs), i.e., AVs with autonomy level 5, are expected 
      to dominate road transportation in the near-future and contribute trillions of dollars to the global economy. 
      The general public, government organizations, 
      and manufacturers all have significant concern regarding resiliency and safety 
      standards of the autonomous driving system (ADS) of AVs . In this work, we proposed and developed (a) `Kayotee' - a fault 
      injection-based tool to systematically inject faults into software and hardware components of the ADS to assess the safety and reliability 
      of AVs to faults and errors, and (b) an ontology model to characterize errors and safety violations impacting reliability and safety of AVs.  Kayotee is capable of characterizing fault propagation and resiliency at different levels - (a) hardware, (b) software, (c) vehicle dynamics, and (d) traffic resilience.  We used Kayotee to study a proprietary ADS technology built by Nvidia corporation and are currently applying Kayotee to other open-source ADS systems.  
      

  \end{abstract}


\IEEEpeerreviewmaketitle
        \section{Introduction}\label{s:intro}
The safety and reliability of autonomous vehicles (AVs) are significant concerns among all the stakeholders. Our previous work ~\cite{banerjee2018hands}
characterized a California Department of Motor Vehicles (DMV) dataset on reported AV road testing and showed that as many as 36\% of disengagements were caused by computer
system problems and 64\% were due to machine learning problems. AV research has traditionally focused on improving 
machine learning and artificial intelligence models.  However, as these models are deployed at large scale on computing platforms, the focus is to assess 
the resilience and safety features of the compute stack driving the AVs. The effects of faults and errors in the 
hardware (GPUs, CPUs and other
processing units) running the AV software stack
is not well understood. Recent work~\cite{Li2017, Pei2017, salami2018resilience, reagen2018ares} exclusively focus on the resiliency
of deep neural networks (DNNs) to hardware faults and errors without accounting for the inherent resiliency in the software stack. \cite{jha2018avfi, rubaiyat2018experimental} 
study the safety of AVs by injecting sensor-related permanent faults such as Gaussian noise, occlusion, etc.  into publicly available autonomous driving system (such as CARLA~\cite{Dosovitskiy17} and Open Pilot~\cite{openpilot}). However, these autonomous driving system (ADS) are overly simplistic with few sensors and are not representative of a production ADS. Moreover, such studies have limited scope as they cannot characterize error masking and propagation of transient, and permanent faults in the ADS. We believe our work is the first to study the impact of
transient errors, intermittent errors and permanent errors (with some limitations) on AV safety and reliability. 

 Our work focuses on developing a fault injection tool to assess hardware and software resilience and its implication on the safety of ADS. We developed `Kayotee' to inject faults into software and hardware components of the ADS in a closed-loop environment to empirically characterize resilience, safety, and the error propagation and masking properties in the ADS system. The fault models were chosen to emulate the representative transient faults in the hardware and software (by corrupting software state variables) to expose error masking limits. The capabilities of Kayotee are (i) injection of hardware and software faults using fault models bundled in the tool, (ii) selection of fault sites based on software components (sensor inputs, object perception, sensor fusion, planner and controller), hardware components (CPU vs. GPU), and machine learning algorithms (DNN vs. non-DNNs), (iii) creation and execution of multiple traffic scenarios, (iv) simulation of the closed-loop environment where parameters (such as speed, acceleration, distance traveled by the AV and actuation command outputs sent by the ADS) follow truncated normal distributions and capture data from multiple modules to compare fault-free run (i.e., golden run) values, i.e., non-injected run outputs to injected run outputs, and (v) calculation of resilience and safety violation parameters.

In comparison to AVFI~\cite{jha2018avfi}, Kayotee is capable of characterizing error propagation and masking in the ADS using a closed-loop simulation environment and is capable of injecting bit flips directly into GPU and CPU architectural state. Our fault injector was tested on the Nvidia DriveWorks platform, and we plan to extend it to Apollo~\cite{apollo} and CARLA~\cite{Codevilla2018, Dosovitskiy17}.

To help improve the safety and reliability characteristics of ADS, we need to answer the following research questions \textemdash
\begin{itemize}
    \item[\textbf{Q1:}] Which software modules are most vulnerable to faults/errors among different parts of the ADS (such as object perception, path perception, localization and planning)?
    \item[\textbf{Q2:}] Do errors in the ADS platform contribute statistically more degradation in safety and resilience characteristics  than degradation because of inherent data quality and inaccuracies in ML/AI-techniques ?
    \item[\textbf{Q3:}] Is there any statistical relationship between safety/resilience and input characteristics (such as \#vehicles, \#people etc.) given errors? 
    \item[\textbf{Q4:}] Is there any statistical difference in error susceptibility among different computing platforms - CPUs and GPUs?
\end{itemize}

Due to proprietary restrictions, we only share fault injection experiment methodology and do not provide any numerical values/results that can identify key-characteristics of the Nvidia driving platform or its susceptibility to faults. The views, opinions, tools and results contained or described in this article are those of the authors and do not necessarily reflect the official policy or position of  Nvidia Corporation. However, we plan to share results with the community by implementing our tools and techniques for open-source ADS and simulators such as Apollo~\cite{apollo} and Carla~\cite{Dosovitskiy17}.


        \begin{figure*}[ht]
    \centering
    \includegraphics[width=0.98\textwidth]{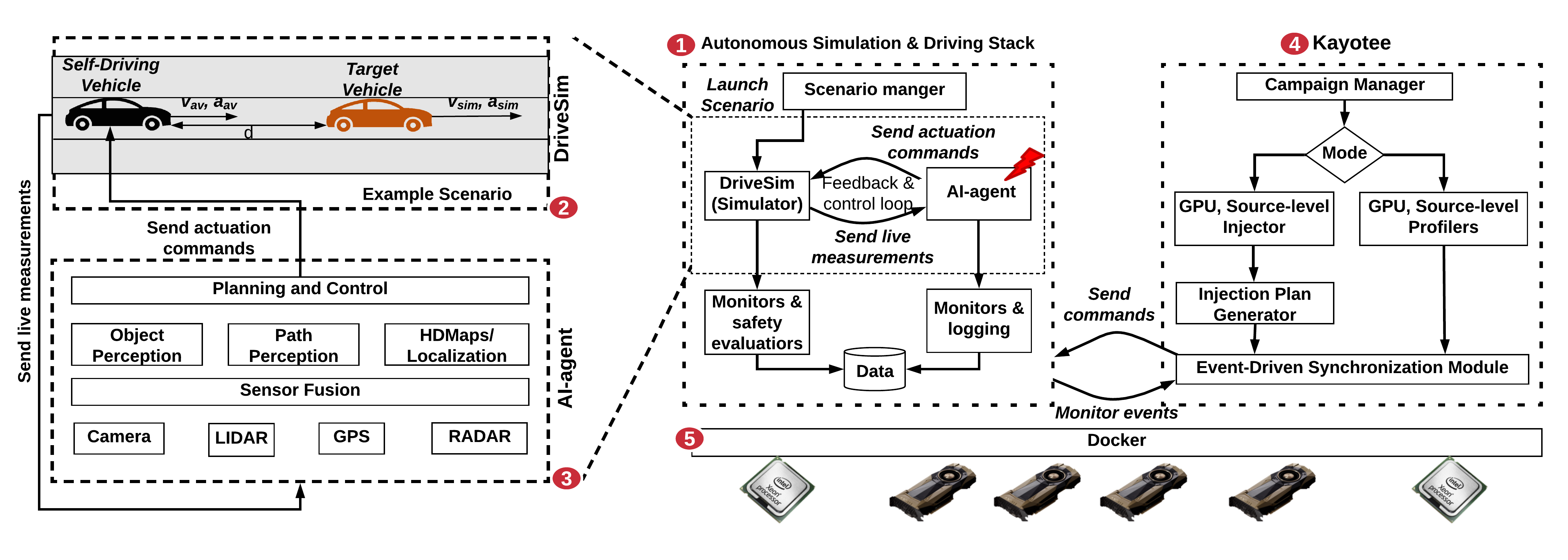}
    \caption{End-to-end safety and reliability evaluation of ADS using Kayotee}
    \label{fig:kayotee}
\end{figure*}
\section{Autonomous Vehicles}
\begin{figure}[t]
    \centering
    \begin{subfigure}[t]{0.45\linewidth}
        \centering
        \includegraphics[width=\linewidth]{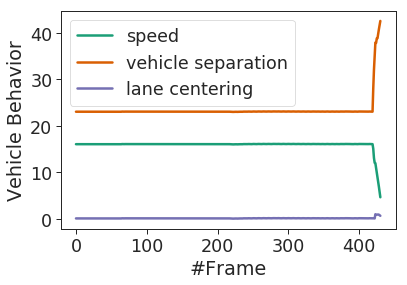}
        \caption{Vehicle behavior}
\label{fig:behavior}
    \end{subfigure}
    \begin{subfigure}[t]{0.47\linewidth}
        \centering
        \includegraphics[width= \linewidth]{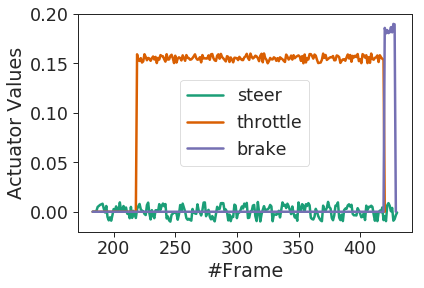}
        \caption{Actuation values}
        \label{fig:actuator}
    \end{subfigure}

    \caption{Simulation results of a single run without fault injection}
    \label{fig:measurements}
\end{figure}

An AV is any vehicle that uses an ADS technology capable of supporting
and replacing a human driver in the tasks of controlling the main functions of steering, acceleration, and monitoring the surrounding environment (e.g., other vehicles/pedestrians, traffic signals, and road markings)~\cite{SAE_J3016_201609}. In this work, we used a proprietary autonomous driving stack (supporting an AI-agent running at automation level 5) and simulation systems (see Fig~\ref{fig:kayotee}, \circled{1}) developed by Nvidia Corporation to showcase the use cases of Kayotee and its usefulness in characterizing the safety and reliability of ADS. Like any other ADS system, the Nvidia ADS consists of the Nvidia AI-agent, marked as \circled{3} in Fig~\ref{fig:kayotee} which further consists of five basic modules - (a) sensors and sensor fusion module, (b) object perception, (c) path perception, (d) maps and localization, and (e) planning and control in addition to several safety check mechanisms (not shown in the figure). In addition to testing Nvidia ADS on private/public roads, Nvidia tests the ADS using a custom Unreal Engine-based simulation engine, DriveSim~\cite{drivesim} (marked as \circled{2}). DriveSim is capable of simulating complex urban scenarios by using a library of urban layouts, buildings, pedestrians, vehicles, and weather conditions (e.g., sunny, rainy, and foggy). An example scenario is shown in \circled{2}. In this scenario, an Nvidia AI-agent controlled vehicle and a DriveSim controlled vehicle are placed on highway driving at different speeds ($v$) and acceleration ($a$) separated by a distance $d$ along with other highway objects (such as road signs, traffic lights, etc. not shown in the figure). Such parameterization allows  mimicking situations such as - (a) a target vehicle slowing down, (b) a stationary target vehicle, (c) an accelerating target vehicle, and many more. The \textit{scenario manager} toolkit in DriveSim can be used to select various pre-created urban scenarios. The \textit{monitoring and safety evaluators} toolkit is used to subscribe to DriveSim measurements providing ground truth values associated with the scenario in realtime and then used to evaluate safety parameters associated with the AI-driven vehicle (such as whether the vehicle is at the center of the lane, whether the vehicle maintains a minimum distance from other vehicles, and whether the speed of the vehicle is within the safety limits with respect to other vehicles and traffic rules). Sample vehicle behavior  (speed, lane centering, and vehicle separation) and actuation output measurements are shown in Fig.~\ref{fig:behavior} and Fig.~\ref{fig:actuator} respectively.


\section{Kayotee Software Architecture}
Kayotee was used to profile the ADS workload running on a computing platform (consisting of Intel Xeon CPUs and Nvidia discrete GPUs\footnote{In this work, we did not inject faults into the Drive AGX Pegasus~\cite{pegasus} platform.}) and then inject faults using representative fault models (by uniform random sampling of fault locations in the GPU architectural state). 

\subsection{Experimental Strategy}\label{sec:strategy}
\begin{figure}
    \centering
    \includegraphics[width=0.45\textwidth]{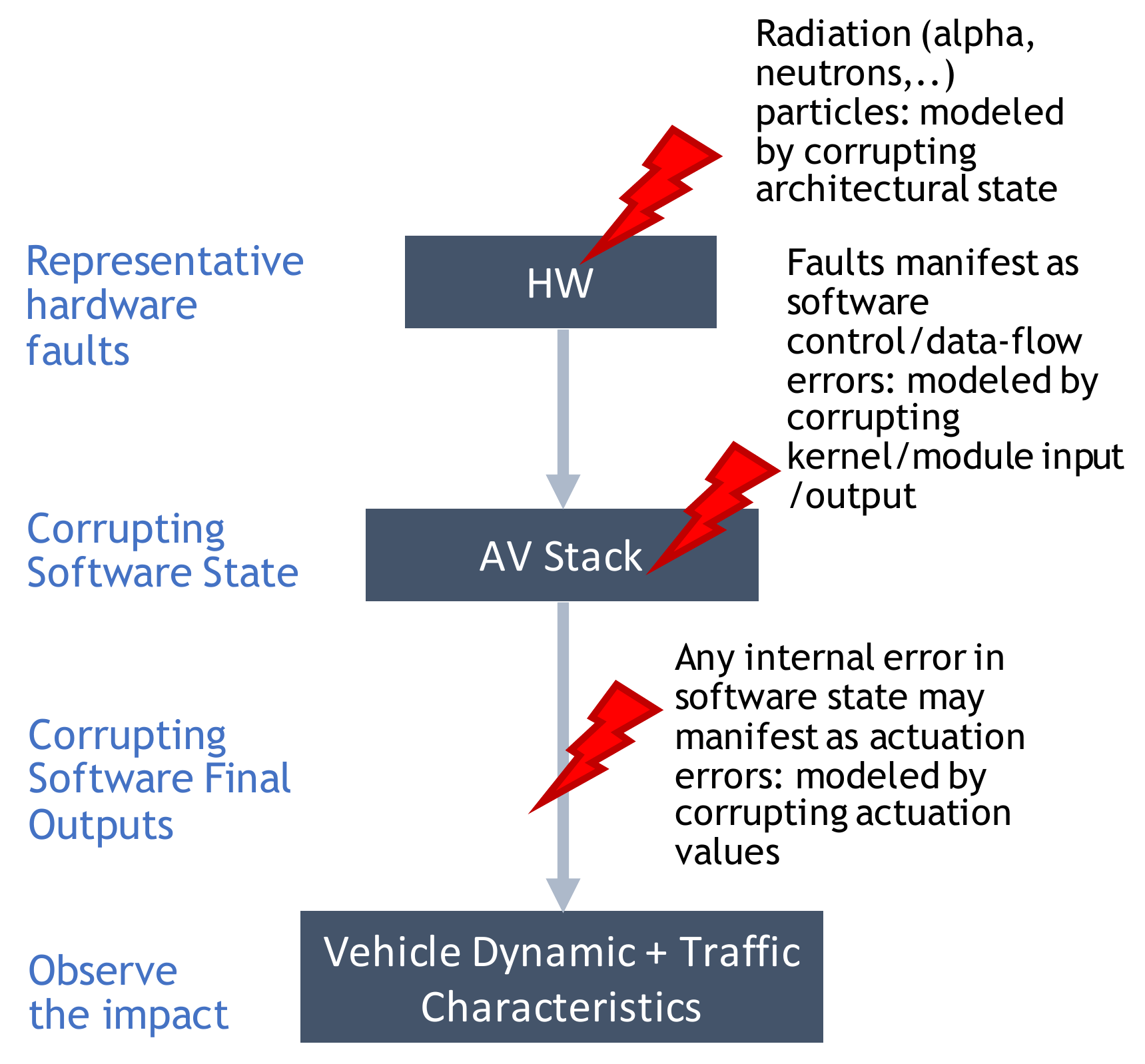}
    \caption{Experimental Strategy}
    \label{fig:strategy}
\end{figure}

We built `Kayotee' to characterize error propagation and masking (a) in the Nvidia GPUs and CPUs, (b) in the ADS, and (c) in vehicle dynamics and traffic.  For each of these characterizations, we built a corresponding injector capable of injecting faults such that errors manifest in the corresponding locations. In the case of GPUs, we used the GPU injector to inject architectural-state faults (see Section~\ref{subsec:gpu_fm}), and SLI (Section~\ref{subsec:sli_fm}) to inject into the inputs and outputs of the ADS kernels (or modules). Corrupting the final output (actuation commands) of the ADS sent to the AV helps us measure resilience associated with vehicle dynamics and traffic. As shown in Figure~\ref{fig:strategy}, low-level circuit-, micro-architectural-, and RTL-faults manifest as architectural-state faults (injected through GPU-injector).  The architectural-state faults that do not get masked manifest as errors in the internal state of the kernels of software stack and any error that does not get masked in the kernel propagates to the output of the kernel. Finally, errors that are not masked before the point of sending actuation commands lead to incorrect actuation commands (actuation errors) sent to the AV. Thus, our approach aids the measurement of fault masking and propagation at different levels and its corresponding impact on the safety of the AV.

\subsection{Kayotee}
Kayotee is a fault injection tool that can inject transient (a)  hardware faults, and (b) software state errors. Kayotee is bundled with a campaign manager that takes an XML configuration file as input to select a fault model, software or hardware module sites for fault injection, the number of faults, and a scenario. The campaign manager uses the specified configuration to (a) profile the ADS workload, (b) generate a fault plan, and (c) inject one or more transient faults per run into the ADS system (until required confidence levels are reached). To monitor and understand the impact of transient faults on the safety and reliability of the ADS, it is important to ensure that the simulation is deterministic.  However, a control-loop based system by definition is non-deterministic in nature. Thus, we developed an `event-driven synchronization' module that coordinated between all the toolkits (DriveSim, monitoring, evaluators, and AI-agent) to ensure that the parameters of the moving objects (AI-vehicle, other vehicles, etc.) in the scenario (including the AI-vehicle) roughly follow a truncated normal distribution. 

\subsection{GPU Injector Fault Models}\label{subsec:gpu_fm}
We consider transient faults in the functional units (e.g., arithmetic and logic units, and load store units), latches, and unprotected SRAM structures of the GPU processor. Such transient faults are modeled by injecting bit-flips (single and double) in the outputs of the executing instructions. If the destination register is a general-purpose register or a condition code, one bit (or two bits) is randomly selected to be flipped. For  store  instructions,  we  flip  a  randomly
selected bit(s) in the stored value. Since  we  inject  errors  directly  into  live  state  (destination
registers), our error model does not account for various masking  factors  in  the  lower  layers  of  the  hardware  stack  such
as circuit-, gate-, and micro-architecture-level masking as well
as masking due to errors in architecturally untouched values. The GPU injection tool uses the same profiling and fault-injection plan generation mechanism as used in SASSIFI~\cite{hari2017sassifi}.
We do not consider faults in cache, memory, and register files because we assume that they are protected by ECC.

\subsection{SLI Fault Models}\label{subsec:sli_fm}
The goal of SLI (Source-Level Injection) is to corrupt the internal state (by corrupting output variables) of the ADS software components. In Table~\ref{tab:sli_fm}, we show some of the variables from each of the ADS modules (see \circled{3} in Figure~\ref{fig:kayotee}) that were targeted using SLI. The fault models supported by SLI are as follows \textendash
\begin{itemize}
    \item \textit{Random}: The chosen variable is randomly modified to a value within the range of zero to vehicle speed limit on the road (e.g., 65mph for majority of US urban interstate roads). For example, in case of object\_class we use possible object classes supported in the Nvidia ADS but for pid\_measured\_values we choose a random value between zero and 65mph (highway speed limit). 
    \item \textit{Fixed}: The chosen variable is always set to the fixed value. It helps to evaluate the worst possible fault cases, e.g., pid\_output for the speed controller is always set to the maximum supported value. This fault-model is useful to inject known fault conditions, and most importantly when generating worse-case intermittent or permanent errors to understand the maximum resiliency offered by the ADS.
    \item \textit{Scale}: The chosen variable is scaled to some ratio of the current value of the variable.
    \item \textit{Disappear}: The chosen output is either not delivered to the next module, if it does not lead to the software crash, or set to null (or zero) if it does lead to software crash.
\end{itemize}
\begin{table}[h]
\centering
\begin{tabular}{|l|p{3.2cm}|p{1.8cm}|}
\hline
\textbf{ADS Module} & \textbf{Output Variables (fault injection target)}                                  & \multicolumn{1}{l|}{\textbf{Fault Model}}                                                          \\ \hline
Path perception     & lane\_type, lane-coordinates                               & random, fixed,                                                                                     \\ \cline{1-2}
Object perception   & num\_detected\_objects, object\_class, object\_coordinates & scale, disappear                                                                                   \\ \cline{1-2}
Planning \& control & pid\_measured\_value, pid\_target\_value, pid\_output      &                                                                                                    \\ \hline
Sensor outputs       & camera frames                                              & Gaussian noise, occlusion (from ~\cite{jha2018avfi}) \\ \hline
\end{tabular}
\caption{SLI supported fault models. Only select few output variables are shown in the table.}
\label{tab:sli_fm}
\end{table}

\subsection{Error and Safety Metrics}
\begin{figure}
    \centering
    \includegraphics[width=0.38\textwidth]{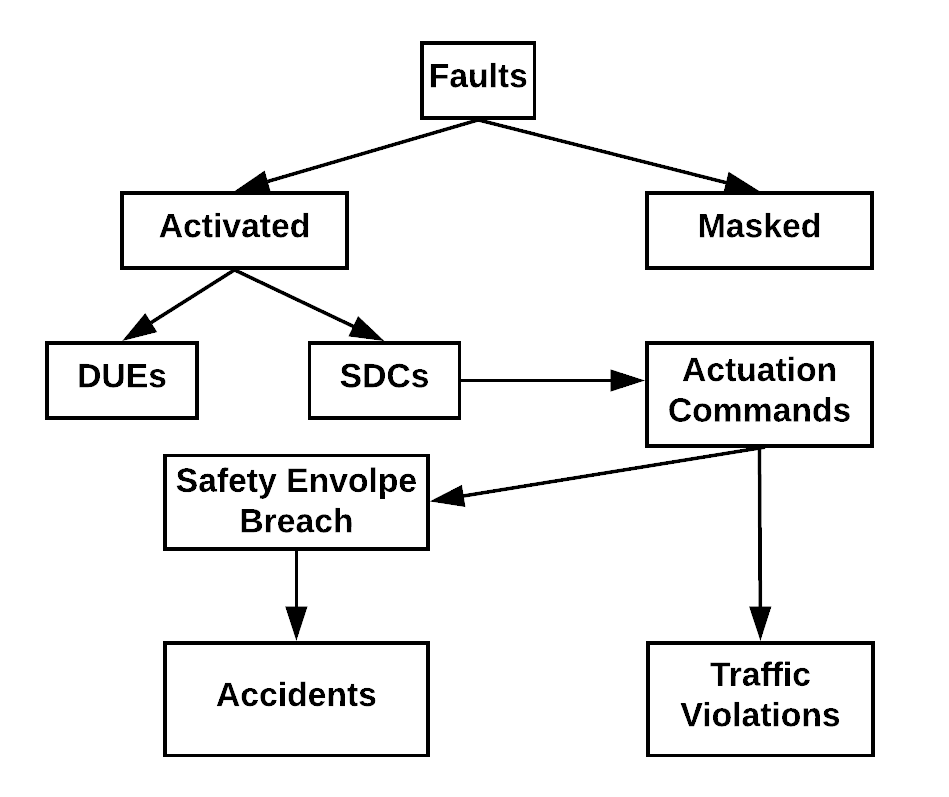}
    \caption{Ontology model for fault manifestation in AVs}
    \label{fig:av-fault}
\end{figure}

To characterize error and safety in the ADS, we propose a new ontology model to capture a range of issues that can occur in the real-world and also observed from fault injection campaigns.  An ontology model capturing the fault manifestation in the ADS is shown in Fig.~\ref{fig:av-fault}. Any run with an injected fault is labeled as \textit{activated} if any of the monitored values (such as object classification, bounding box, actuator command values, vehicle measurements, etc.) do not fall within the expected value range obtained in the golden runs, i.e., a run with no injection or if leads to a hang or crash. We use an IQR (interquartile range)-based outlier detection algorithm~\cite{analytical1989robust} combined with the range of the distribution in the golden run to label a variable as containing an erroneous value. Such error labeling is scenario-specific as the ground truth values are obtained by running the ADS for a specific scenario with no injection;  otherwise, it is labeled as \textit{masked}. An activated fault can be further classified as \textit{DUE} (detectable uncorrectable error, such as a hang or crash) or \textit{SDC} (silent data corruption). We label a run as SDC if there is a change in the value of any of the monitored variables in the AV stack compared to the golden run value. SDCs can lead to actuation errors, and any such run is labeled as \textit{actuation-error}.  Actuation errors can lead to a breach in the safety envelope or traffic violations or both. A breach in the safety envelope can lead to an accident.

In our study, we defined following scenario-agnostic safety metrics for breaches of safety:

\textbf{Safety envelope breach} occurs when the collision distance ( i.e., which is the distance traveled by the vehicle from its current position to collision point) between AV and any other object on the road (moving or stationary) is less than the stopping distance (refer to section 2.6.1 of ~\cite{cadmv}). The collision point can be calculated using trajectory estimation approaches~\cite{brown2017safe}.  The stopping distance ($D_s$) is given by 
    \begin{equation}
         D_s = D_p + D_r + D_b
    \end{equation}
    \textit{Perception distance ($D_p$)} is the distance the vehicle travels in ideal conditions from the time that the driver (human or AI-agent) of the vehicle sees a hazard until the brain or ADS recognizes it. The average perception time for an alert human driver is 1.75 seconds~\cite{cadmv}. For an AI-agent the worst-case recognition time is $1/{FPS}$, where FPS is the frames per second processed by the AI-agent. 
    
    \textit{Recognition distance ($D_r$)} is the distance the vehicle will continue to travel in ideal conditions before physically hitting the brakes in response to a hazard seen ahead. The average human driver has a reaction time of 0.75 to 1.0 seconds~\cite{cadmv}. For an AI-agent, this corresponds to the time taken to send an actuation command after recognizing the hazard. 
    
    \textit{Braking distance ($D_b$)} is the distance the vehicle will travel in ideal conditions while braking. On a highway at 24.5872 meters/sec (55mph), a vehicle will travel a minimum of 64 meters of braking distance~\cite{cadmv}. If two vehicles are on a highway traveling in the same direction, the vehicle separation may only be 20 meters, but the collision distance is significantly higher due to vehicle dynamics. The braking distance depends on the road surface and the type, weight, speed, and acceleration of the vehicle.  In this paper, we only considered speed to calculate braking distance.  However, the braking distance can be more accurately calculated using previously proposed models~\cite{wu2010test, yi2003emergency}.

\textbf{Lane centering breach} occurs when the distance from the lane center changes by more than 0.5 meters. 


        \balance

        \section{Conclusion and Future Work}\label{sec:conclusion}

In this work we presented Kayotee, a fault injection tool, along with methodologies to empirically assess the fault propagation, resilience, and safety characteristics of the ADS. The main objective of Kayotee is to evaluate the effects of the faults on the ADS and to investigate the maximum number of the faults that the ADS can tolerate before a safety violation occurs.  Although Kayotee may help identify some of the software design issues or bugs in the ADS, it does not systematically evaluate the software design or bugs (e.g., using static checkers). 

The future work involves \textemdash 
\begin{itemize}
   \item porting Kayotee to the publicly available open-source Apollo~\cite{apollo} system and to share the resilience/safety characteristics with the community, 
   \item beam-testing of the CPUs and GPUs to understand the differences in emulated faults on CPUs/GPUs to representative faults that may occur in the field.
\end{itemize}

\section*{Acknowledgments} \addcontentsline{toc}{section}{Acknowledgment}
This material is based upon work supported by an IBM Faculty Award, and by the National
Science Foundation (NSF) under Grant Nos. ACI 1535070 and CNS 15-45069.
Any opinions, findings, and conclusions or recommendations expressed in this material
are those of the authors and do not necessarily reflect the views of the NSF. We thank K. Atchley, Ankit Gupta , Vibhor Rastogi, and Matt Campbell for their help.

\bibliographystyle{IEEEtran}
\bibliography{./references}
\end{document}